\documentclass[preprint,aps,12pt,showpacs,nofootinbib,tightenlines]{revtex4}
\usepackage{amsmath}
\usepackage{amssymb}
\usepackage{epsfig}
\usepackage{graphicx}
\begin{document}

%%%%%%%%%%%%%%%%%%%%%%%%%%%%%%%%%%%%%%%%%%%%%
%%%%%%%%%%%%%%%%%%%%%%%%%%%%%%%%%%%%%%%%%%%%%%%%%%%%%%
\def\pslash{\rlap{\hspace{0.02cm}/}{p}}
\def\eslash{\rlap{\hspace{0.02cm}/}{e}}
\def\Eslash{\rlap{\hspace{0.02cm}/}{E}}
%%%%%%%%%%%%%%%%%%%%%%%%%%%%%%%%%%%%%%%%%%%%%%%%%%%%%%
\title{The flavor-changing bottom-strange quark production in the
littlest Higgs model with T parity at the ILC}

\author{ Bingzhong Li$^{1,2}$}\email{libingzhong08@yahoo.cn}
\author{ Jinzhong Han $^1$}\email{hanjinzhongxx@126.com}
\author{ Bingfang Yang $^1$}

\affiliation{
$^1$ College of Physics and Information Engineering, Henan Normal University, Xinxiang, Henan 453007, P.R.China \\
$^2$ College of Physics and Electronic Engineering, Xinyang Normal
     University, Xinyang, Henan 464000, P.R.China \vspace*{1.5cm}}

\begin{abstract}

In the littlest Higgs model with T-parity (LHT) the mirror quarks
induce the special flavor structures and some new flavor-changing
(FC) couplings which  could  greatly enhance the production rates of
the FC processes. We in this paper study some bottom and
anti-strange production processes in the LHT model at the
International Linear Collider (ILC), i.e., $e^+e^-\rightarrow
b\bar{s}$ and $\gamma\gamma\rightarrow b\bar{s}$. The results show
that the production rates of these processes are sizeable for the
favorable values of the parameters. Therefore, it is quite possible
to test the LHT model or make some constrains on the relevant
parameters of the LHT through the detection of these processes at
the ILC.

\end{abstract}


\maketitle

%%%%%%%%%%%%%%%%%%%%%%%%%%%%%%%%%%%%%%%%%%%%%%%%%%%%%%%%%%%%%%%%%%%%%%%%%%%%%%%
\section{ Introduction}

The Standard Model (SM) suffers from shortcomings, such as the
hierarchy problem, little Higgs model offers an alternative route to
solve this problem \cite{little Higgs}. The littlest Higgs (LH)
model is the most economical implementation of the little Higgs idea
\cite{LH}, but which is found to be subject to strong constraints
from electroweak precision tests \cite{constraints}. In order to
evade such restrictions on the LH model, one of the most attractive
models is proposed which is just the Littlest Higgs model with
T-parity (LHT) \cite{LHT}, where the discrete symmetry named
``T-parity" forbids tree-level corrections to electroweak
observables, thus weakening the electroweak precision constraints.
In the LHT model, the flavor structure is richer than the that of
the SM, mainly due to the presence of three doublets of mirror
quarks and leptons that interact with the ordinary quarks and
leptons. The appearance of the new flavor-changing (FC) interactions
in the LHT model can have crucial phenomenology at the high energy
colleders and studied in Ref. \cite{yue-LHT}. As we know, the SM
does not exist the tree-level FC neutral currents, though it can
occur at higher order through radiative corrections, and these
effects are hardly to be observed because of the loop suppression.
The new FC interactions in the LHT model could significantly enhance
the FC processes which can make some FC processes observable. Hence
the FC processes provide an excellent opportunity to probe new
physics due to their clean SM backgrounds.

The FC interactions in the LHT model could induce the loop-level
$t\bar{c}V$ and $b\bar{s}V~(V=\gamma, Z,g)$ couplings. The
$t\bar{c}V$ coupling can contribute to the rare top quark decays
$t\rightarrow cV~(V=\gamma, Z,g)$ \cite{tcv-LHT,yang-LHT} and some
FC production processes \cite{yang-LHT,eeppeq-LHT}. On the other
hand, Ref. \cite{zbs-LHT} has performed a collective study for the
various FC decays of $B$-meson, $Z$-boson and Higgs boson, and it is
found that the LHT predictions obviously deviate from the SM
predictions. The final states of $b\bar{s}$ can be produced via the
FC interactions. At the linear collider, the production process
$e^+e^-\rightarrow b\bar{s}$ has been investigated in the SM
\cite{eebs-SM},  the topcolor-assistant multiscale technicolor model
\cite{eebs-TC2}, and the the 4 generation Standard Model
\cite{SM-Four}. In addition, the production process  $pp\rightarrow
h\rightarrow b\bar{s}
 (h =h^0 ,H^0 ,A^0)$ was also discussed in the Minimal
Supersymmetric Standard Model (MSSM) at the large hadron collider
(LHC) \cite{eebs-MSSM}. Therefore, it is worthy to probe the
production processes
 $e^+e^-(\gamma\gamma)\rightarrow b\bar{s}$ in the LHT model
 at the international linear collider (ILC) .

Although the LHC will essentially enlarge the possibilities of
testing for new physics effects,  the analysis of new physics
processes at LHC is complicated. An $e^+e^-$ facility with clean
environments (and, potentially, with various options such as $\gamma
e$ and $\gamma\gamma$ collision modes) is required to complement
this hadron machine in drawing a comprehensive and high-resolution
picture of electroweak symmetry breaking and of the new physics
beyond the SM. The ILC \cite{ILC}, which is currently being
designed, would be an excellent counterpart to LHC.

\indent  This paper is organized as follows. In Sec. II, we briefly
review the LHT model. Sec. III presents the detailed calculation of
the production cross sections of the processes. The numerical
results are shown in Sec. IV. Finally, a short conclusion was given
in Sec. V.

\section{  A brief review of the LHT model }
 The original LH model\cite{LH}, which is based on a non-linear sigma
  model, begins with $SU(5)$ global symmetry, with a locally gauged subgroup
  $G_1\otimes G_2=[SU(2)_1\otimes U(1)_1]\otimes[SU(2)_2\otimes U(1)_2]$.
  With the elegant breaking mode of the global symmetry group $SU(5)$
  and the special structure of the gauge symmetry group $G_1\otimes G_2$,
  the discrete symmetry group $\mathbb{Z}_2$ dubbed ``T-parity" can be facilely implemented in the
  LH model, which makes the explicit group structures of the LHT model.

  From the $SU(5)/SO(5)$  breaking at the scale $f\sim\mathcal {O}(TeV)$,
  there arise 14 Nambu-Goldstone bosons which
  are described by the ``pion" matrix $\Pi$, given explicitly by
\begin {equation}
\Pi=
\begin{pmatrix}
-\frac{\omega^0}{2}-\frac{\eta}{\sqrt{20}}&-\frac{\omega^+}{\sqrt{2}}
&-i\frac{\pi^+}{\sqrt{2}}&-i\phi^{++}&-i\frac{\phi^+}{\sqrt{2}}\\
-\frac{\omega^-}{\sqrt{2}}&\frac{\omega^0}{2}-\frac{\eta}{\sqrt{20}}
&\frac{v+h+i\pi^0}{2}&-i\frac{\phi^+}{\sqrt{2}}&\frac{-i\phi^0+\phi^P}{\sqrt{2}}\\
i\frac{\pi^-}{\sqrt{2}}&\frac{v+h-i\pi^0}{2}&{\sqrt{4/5}}\eta&-i\frac{\pi^+}{\sqrt{2}}&
\frac{v+h+i\pi^0}{2}\\
i\phi^{--}&i\frac{\phi^-}{\sqrt{2}}&i\frac{\pi^-}{\sqrt{2}}&
-\frac{\omega^0}{2}-\frac{\eta}{\sqrt{20}}&-\frac{\omega^-}{\sqrt{2}}\\
i\frac{\phi^-}{\sqrt{2}}&\frac{i\phi^0+\phi^P}{\sqrt{2}}&\frac{v+h-i\pi^0}{2}&-\frac{\omega^+}{\sqrt{2}}&
\frac{\omega^0}{2}-\frac{\eta}{\sqrt{20}}
\end{pmatrix}
\end{equation}
The fields $\omega^0, \omega^\pm$ and $\eta$ are Goldstone bosons
eaten by new T-odd heavy gauge bosons $W_H$, $Z_H$ and $A_H$, which
obtain masses at $\mathcal {O}(v^2/f^2)$
\begin{equation}
M_{W_{H}^{\pm
}}=M_{Z_{H}}=gf(1-\frac{v^2}{8f^2}),~~~~~~~~M_{A_H}=\frac{g'}{\sqrt{5}}f(1-\frac{5v^2}{8f^2})
\end{equation}
  where $g$ and $g'$ are the SM $SU(2)$ and $U(1)$ gauge couplings,
respectively.

   According to the T-parity in the LHT model,
  the T-even fermion section consists of the SM quarks,
  leptons and a color triplet heavy quark $T_{+}$, and the T-odd fermion sector
  consists first of all of three generations of mirror quarks and leptons
  with vectorial couplings $SU(2)_L$ and a T-odd heavy top quark $T_-$.
  In this paper only mirror quarks are relevant which are denoted by

\begin{eqnarray}
   \left( \begin{array}{c} u^1_H\\d^1_H \end{array} \right)~,~~~~
   \left( \begin{array}{c} u^2_H\\d^2_H \end{array} \right)~,~~~~
   \left(\begin{array}{c} u^3_H\\d^3_H \end{array} \right)
\end{eqnarray}
  with their masses satisfying to first order in $\mathcal {O}(v/f)$
\begin{eqnarray}
     m_{u_H}^{1}=m_{d_H}^{1},~~~~~~m_{u_H}^{2}=m_{d_H}^{2},~~~~~~m_{u_H}^{3}=m_{d_H}^{3}.
\end{eqnarray}

  For completeness we would like to mention that the SM electroweak gauge bosons ($W^{\pm}_L,Z_L,A_L$)
  are belong to the T-even sector and a Higgs triplet $\Phi$ belongs
  to the T-odd sector. The charged Higgs $\phi^{\pm}$, as well as the neutral
  Higgses $\phi^{0}$, $\phi^{P}$, are relevant in principle for the production processes
  considerated  in our paper, but their effects turn out to be of
  high order in $\upsilon/f$ \cite{FC-LHT1}, and consequently
  similarly  to $T_+$ will not enter our analysis.

  One of the most important ingredients of the mirror quark sector
  is the existence of the flavor violating actions between
  the SM fermions and the mirror fermions which are mediated by
  the T-odd heavy gauge bosons, which leads to the appearance of
  two CKM-like unitary mixing matrices $V_{H_u}$ and $V_{H_d}$, that satisfy
\begin{eqnarray}
V_{H_u}^{\dagger}V_{H_d}=V_{CKM}
\end{eqnarray}
   The notation indicates the type of light fermion that is involved
   in the interaction, i.e. if it is of up- or down-type.

 Following \cite{LHT,FC-LHT2,cp-yun} we will parameterize $V_{H_d}$ generalizing
 the usual CKM parameterisation, as a product of three rotations,
 and introducing a complex phase in each of them , thus obtaining
\begin{eqnarray}
V_{H_d}=
\begin{pmatrix}
c^d_{12}c^d_{13}&s^d_{12}c^d_{13}e^{-i\delta^d_{12}}&s^d_{13}e^{-i\delta^d_{13}}\\
-s^d_{12}c^d_{23}e^{i\delta^d_{12}}-c^d_{12}s^d_{23}s^d_{13}e^{i(\delta^d_{13}-\delta^d_{23})}&
c^d_{12}c^d_{23}-s^d_{12}s^d_{23}s^d_{13}e^{i(\delta^d_{13}-\delta^d_{12}-\delta^d_{23})}&
s^d_{23}c^d_{13}e^{-i\delta^d_{23}}\\
s^d_{12}s^d_{23}e^{i(\delta^d_{12}+\delta^d_{23})}-c^d_{12}c^d_{23}s^d_{13}e^{i\delta^d_{13}}&
-c^d_{12}s^d_{23}e^{i\delta^d_{23}}-s^d_{12}c^d_{23}s^d_{13}e^{i(\delta^d_{13}-\delta^d_{12})}&
c^d_{23}c^d_{13}
\end{pmatrix}
\end{eqnarray}
  As in the case of the CKM matrix the angles $\theta^d_{ij}$ can
  all be made lie in the first quadrant with
  $0\leq\delta^d_{12},\delta^d_{23},\delta^d_{13}\leq 2\pi$.
  The matrix $V_{H_u}$ is then determined through
  $V_{H_u}=V_{H_d}V_{CKM}^{\dagger}$.

\section{The bottom-strange quarks production in the LHT model at the ILC}

 In the LHT model, the Feynman diagrams of the process $e^+e^- \to
b\bar{s}$ are shown in Fig.1 (a) and the Feynman diagrams of the
process $\gamma\gamma \to b\bar{s}$ are shown in Fig.1 (b) and (c).
The black square in Fig. 1 denotes the effective FC couplings
$b\bar{s}\gamma(Z) $ which were presented in Fig.2.
 As we have mentioned above,
 there are FC interactions between SM fermions and mirror fermions
 which are mediated by the heavy gauge bosons ($A_H,Z_H,W^{\pm}_H$)
 or Goldstone bosons ($\eta,\omega^0,\omega^{\pm}$) under 't Hooft-Feynman gauge.
 Therefore, we also plot the diagrams involving
 the Goldstone bosons $\eta$, $\omega^0$ and $\omega^{\pm}$
 in Fig.2.
 In addition, the Goldstone boson mass is conveniently taken the same of its corresponding gauge
 boson, i.e. $m_{\eta}=M_{A_{H}}$ and
 $m_{{\omega}^{0,\pm}}=M_{W_{H}}$.
 With the FC couplings given by Ref. \cite{FC-LHT1}, the
 loop-level FC couplings $b\bar{s}Z(\gamma)$ can be induced.

\begin{figure}
%\begin{center}
\scalebox{0.5}{\epsfig{file=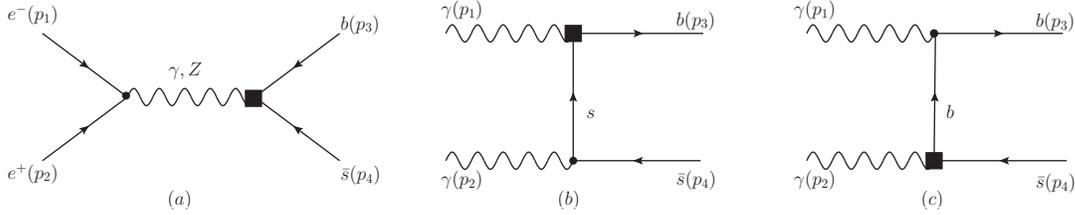}} \caption{(a): The Feynman
diagrams of the process $e^+e^-\to b\bar{s}$. (b) and (c): Diagrams
of the process $\gamma\gamma\to b\bar{s}$.  The diagrams with the
interchange of the two incoming photons are not shown here.}
%\end{center}
\end{figure}

 One important difference of the $b\bar{s}\gamma $
vertex in $e^+e^-\rightarrow b\bar{s}$ and in $\gamma \gamma
\rightarrow b\bar{s}$ is both the quarks are on-shell for
$e^+e^-\rightarrow b\bar{s}$, while for $\gamma\gamma\rightarrow
b\bar{s}$, either $b$ or $\bar{s}$ is off-shell, which  is almost
effortless to find from Fig.1. In order to simplify calculation,
we'd better obtain an universal form of the
 $b\bar{s}\gamma$ vertex which can be applied to the both cases,
 and the best possible way was suggested by \cite{effective vertex}. In our
calculation, we use the method in \cite{effective vertex} to get the
effective $b\bar{s}\gamma(Z)$ vertex $\Gamma^{\mu}_{b\bar{s}\gamma}$
and $\Gamma^{\mu}_{b\bar{s}Z}$, which can be directly calculated
based on Fig.2 and represented in form of 2-point and 3-point
standard functions $B_0,B_1,C_{ij}$. Because the analytical
expressions are lengthy and tedious, we will not present them in our
paper. we calculate the amplitudes numerically by using the
LOOPTOOLS \cite{LoopTools}.

\begin{figure}
\begin{center}
\scalebox{0.5}{\epsfig{file=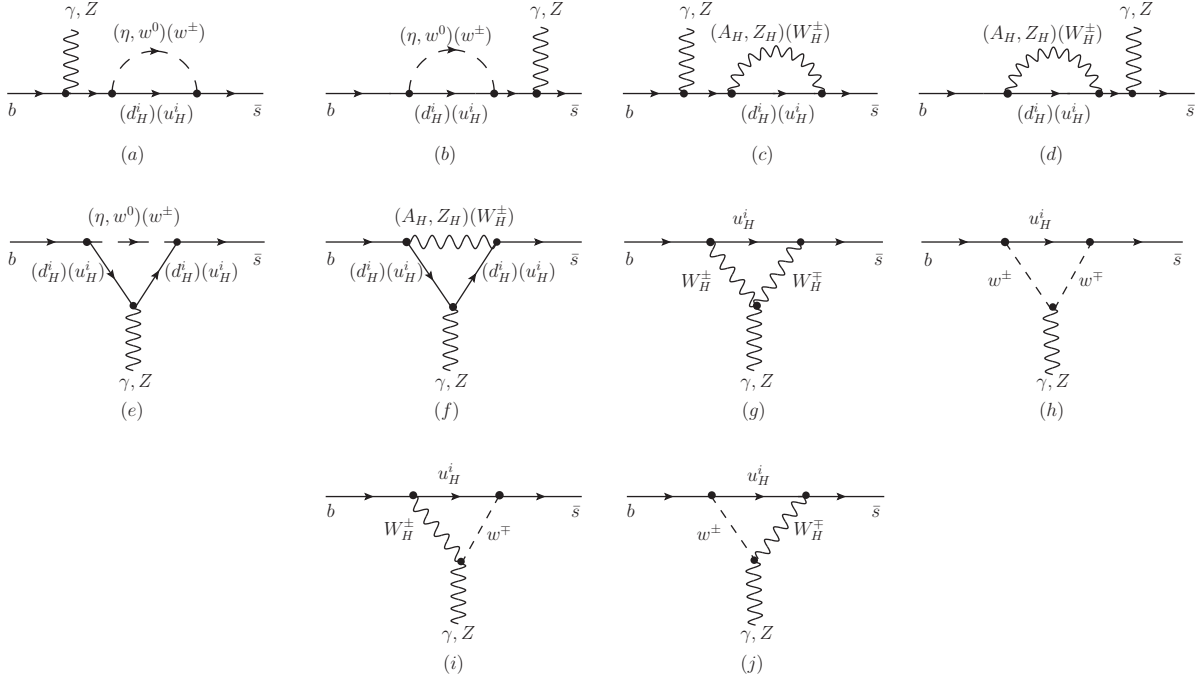}}  \caption{The Feynman
diagrams of the one-loop contributions of the LHT model to the
effective couplings $b\bar{s}\gamma(Z)$.}
\end{center}
\end{figure}

\section{The numerical results of the cross sections for the processes
$e^+e^-(\gamma\gamma)\rightarrow b\bar{s}$ in the LHT model}

There are several free parameters in the LHT model which are
involved in the amplitudes of $e^+e^-(\gamma\gamma)\rightarrow
b\bar{s}$. They are the breaking scale $f$, the masses of the mirror
quarks $m_{H_i}$ (we take $m^i_{u_H}=m^i_{d_H}=m_{H_i},i=1,2,3$, as
in equation (4)) , and 6 parameters
($\theta^d_{12},~\theta^d_{13},~\theta^d_{23},~\delta^d_{12},~\delta^d_{13},~\delta^d_{23}$)
which are related to the mixing matrix $V_{H_d}$. For the parameter
$f$,  we take two discrete typical value in our calculation, i.e.
$f=500$ GeV and 1000 GeV, according to the investigation of
\cite{LHT} which  pointed out that the value of $f$ can be as low as
500 GeV.

In Refs. \cite{FC-LHT2,FC-LHT1}, the constraints on the mass
spectrum of the mirror fermions have been probed in the study of
neutral meson mixing in the $K,~B$ and $D$ systems. They show that a
TeV scale GIM suppression is necessary for a generic choice of
$V_{H_d}$. However, there are regions of parameter space where are
only very loose constraints on the mass spectrum of the mirror
fermions. For the mixing matrix $V_{H_d}$, we take two scenarios for
these parameters according to \cite{zbs-LHT}  and fulfill our
calculation:
\begin{equation}
{\rm Case ~I}: V_{H_u}=I, ~V_{H_d}=V_{\rm
CKM}.~~~~~~~~~~~~~~~~~~~~~~~~~~~~~~~~~~~~~~~~~~~~~~~~~~~~~~~~~~~~~~~~~~~~~~~~~~~~~~~~~~~~~~~~~~\nonumber
\end{equation}
\begin{equation}
{\rm Case ~II}:\delta_{12}^{d}=\delta_{23}^{d}=\delta_{13}^{d}=0,\\
 1/\sqrt{2}\leq
s_{12}^{d}\leq 0.99,5\times 10^{-5}\leq s_{23}^{d}\leq2\times
10^{-4},4\times 10^{-2}\leq s_{13}^{d}\leq 0.6.\nonumber
\end{equation}
To fix matrix $V_{H_d}$ in Case II, we adopt the upper limit and
  down limit of $s^d_{ij}$, respectively,
\begin{equation}
{\rm Case
 ~II(1)}:\delta_{12}^{d}=\delta_{13}^{d}=\delta_{23}^{d}=0,~
s_{12}^{d}=1/\sqrt{2},~ s_{23}^{d}=5\times 10^{-5},~
s_{13}^{d}=4\times 10^{-2}, ~~~~~~~~~~~~~~~~~\nonumber
\end{equation}
\begin{equation}
\hspace{-0.7cm} {\rm Case~
II(2)}:\delta_{12}^{d}=\delta_{13}^{d}=\delta_{23}^{d}=0,~
s_{12}^{d}=0.99,~ s_{23}^{d}=2\times 10^{-4},~ s_{13}^{d}= 0.6.
~~~~~~~~~~~~~~~~~~~ \nonumber
\end{equation}
In both the above cases, the constraints on the mass spectrum of the
mirror fermions are very relaxed. On the other hand, Ref. \cite{EW
constraint} has shown that the experimental bounds on four-fermi
interactions involving SM fields provide an upper bound on the
mirror fermion masses and this yields $m_{H_i}\leq {4.8f^2}/{\rm
TeV}$. Here, we also consider such constraint and let $m_{H_3}$ to
vary in the range of $600-1200$ GeV  and $600-4800$ GeV when $f=500$
GeV and 1000 GeV respectively. Meanwhile, we fix
$m_{H_1}=m_{H_2}$=500 GeV.

 In the numerical calculation, there are also a set of independent input parameters,
 and they are  $m_{b}=$4.2 GeV, $m_{s}=$0.095 GeV,
 $ m_{Z}=91.188$ GeV, $\alpha_e=1/128$, and
 $\sin^2\theta_w=0.231$ \cite{PDG}. For the c.m. energies of the ILC, we choose
$\sqrt{s}=500$ GeV and 1000 GeV according to \cite{ILC}. The final
numerical results are shown in Figs.3-5.

\begin{figure}[htbp]
\includegraphics[width=2.8in,height=2.8in]{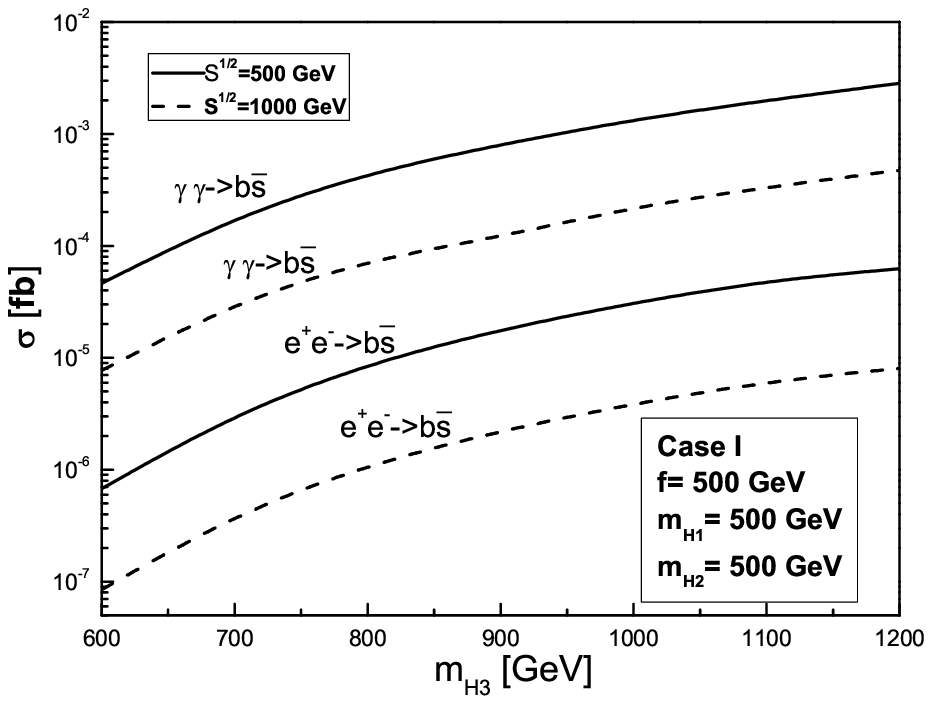}
\hspace{0in}
 \includegraphics[width=2.8in,height=2.8in]{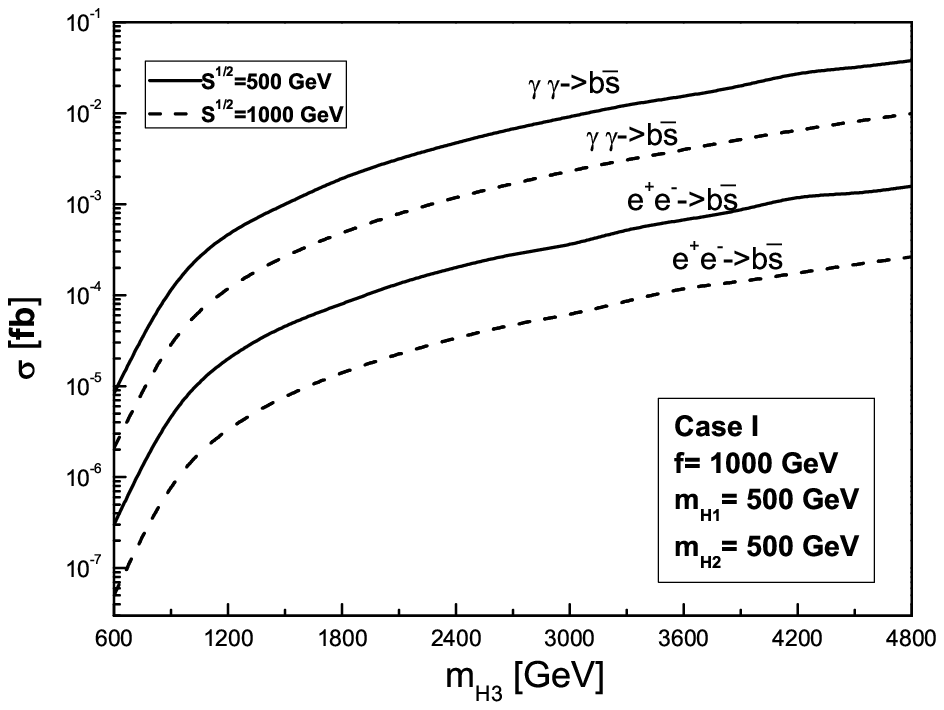}\\
\caption{\small The cross sections of the processes
$e^+e^-(\gamma\gamma)\rightarrow b\bar{s}$ in the LHT model for Case
I, as a function of $m_{H_3}$.}
\end{figure}

\begin{figure}[htbp]
\includegraphics[width=2.8in,height=2.8in]{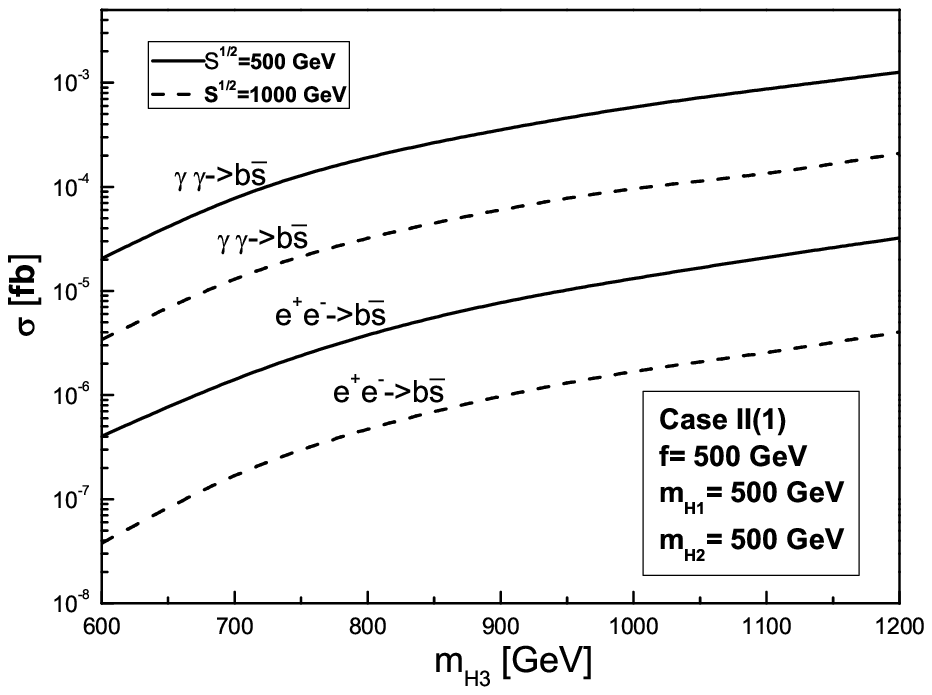}
\hspace{0in}
 \includegraphics[width=2.8in,height=2.8in]{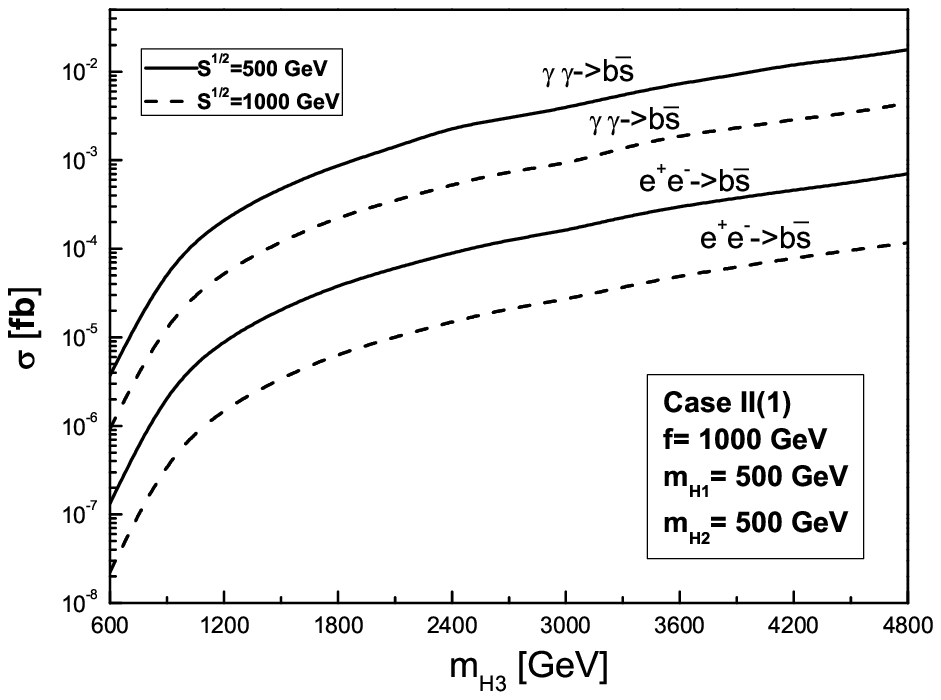}\\
\caption{\small The cross sections of the processes
$e^+e^-(\gamma\gamma) \rightarrow b\bar{s}$ in the LHT model for
Case II(1) , as a function of $m_{H_3}$.}
\end{figure}

\begin{figure}[htbp]
\includegraphics[width=2.8in,height=2.8in]{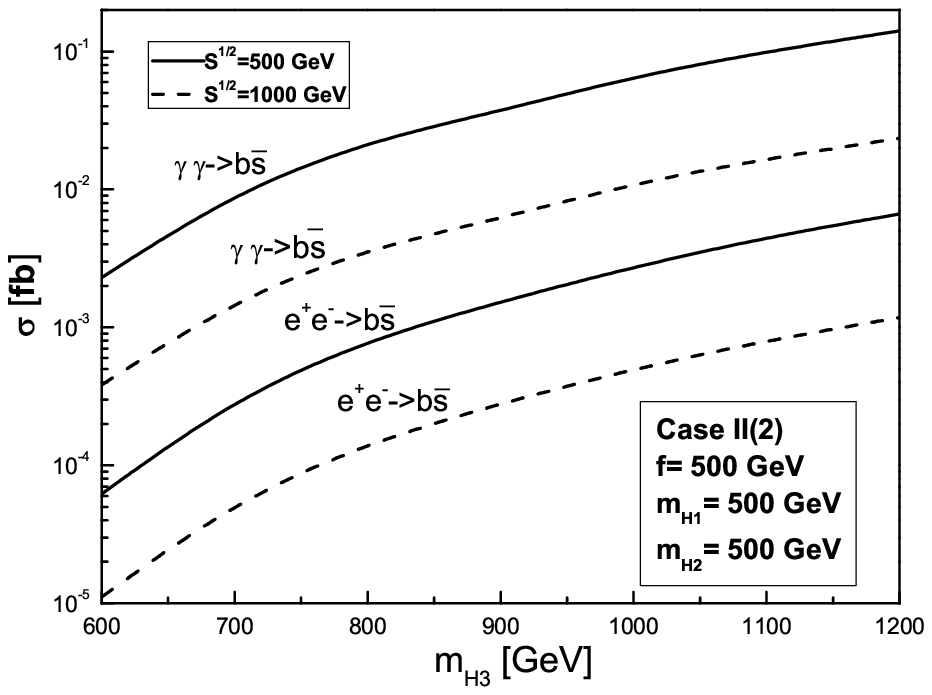}
\hspace{0in}
 \includegraphics[width=2.8in,height=2.8in]{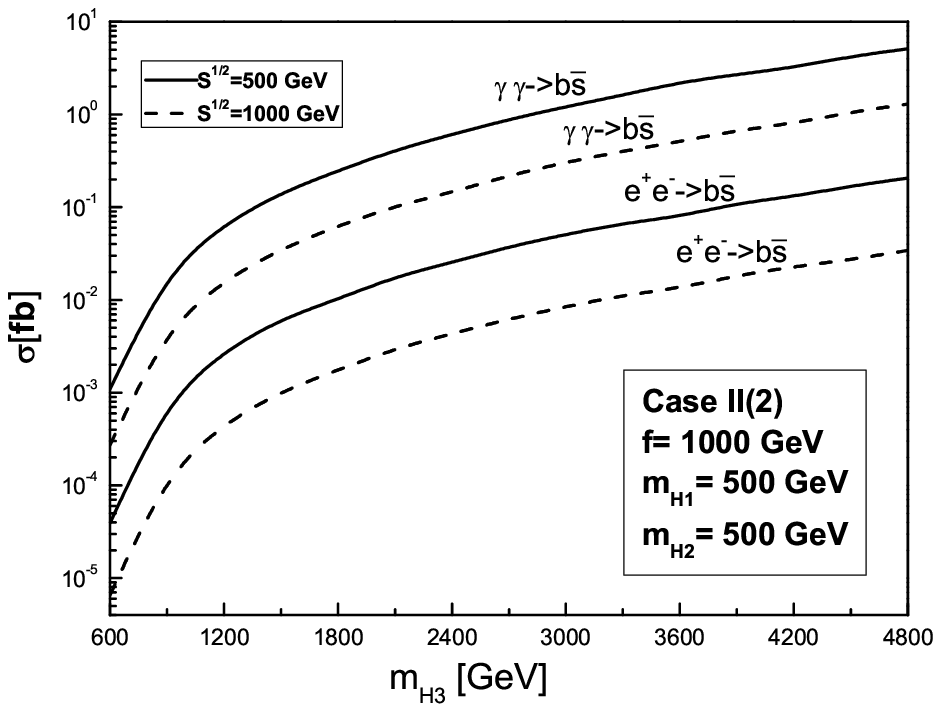}\\
\caption{\small The cross sections of the processes
$e^+e^-(\gamma\gamma) \rightarrow b\bar{s}$ in the LHT model for
Case II(2) , as a function of $m_{H_3}$.}
\end{figure}
In Figs.3-5, we plot the cross sections of the processes
$e^+e^-(\gamma\gamma)\rightarrow b\bar{s}$ as a function of
$m_{H_3}$ for case I, case II(1) and case II(2), respectively, and
other relevant parameters  were presented in the corresponding
figure. From these figures, we can see the following: (i) The cross
sections of the processes become larger with the $m_{H_3}$
increasing. This is because these processes proceed in a way quite
similar to the GIM mechanism of the SM, so the more significant the
mass splitting between the mirror quarks, the larger the rates
become. And hence, through the detection of these production
process, one can set some constrains on the masses of the mirror
quarks. (ii) It is evident that the cross section of
$\gamma\gamma\rightarrow b\bar{s}$ is larger than that of
$e^+e^-\rightarrow b\bar{s}$ in the whole $m_{H_3}$ region, and the
difference between these two cross sections is about one or two
orders of magnitude. In general, this is because the process
$e^+e^-\rightarrow b\bar{s}$ occurs only via s-channel, so its rate
is suppressed by the photon propagator and the $Z$ propagator.
However, the process $\gamma\gamma\rightarrow b\bar{s}$ may overcome
this handicap. Therefore, for the production of $b\bar{s}$, the
$\gamma\gamma$ collider has more advantage than the $e^+ e^-$
collider in the LHT model. (iii) The figures illustrate that the
cross sections of the processes $e^+e^-(\gamma\gamma)\rightarrow
b\bar{s}$ are quite different for the different schemes of the
parameterization of the mixing matrix $V_{H_d}$, especially for case
II(2) illustrated in Fig.3 and case I illustrated in  Fig.5 , which
demonstrate that the cross sections are strongly dependent on the
parameterization scheme of $V_{H_d}$. The possible reason for such
difference is the parameters configuration of the mixing matrix
$V_{H_d}$, and such difference could help us to confirm the flavor
structure of the LHT model and then to validate the parameterization
of the mixing matrix $V_{H_d}$ if these production processes could
be detected at the ILC. (iv) The cross sections of the processes
$e^+e^-(\gamma\gamma)\rightarrow b\bar{s}$ are insensitive to the
scale $f$. The reason is that the masses of the heavy gauge bosons $
M_{V_H}$ and the mirror quarks $m_{H_i}$ are proportional to $f$,
but the production amplitudes are represented in the form of
$m_{H_i} /M_{V_H}$ which cancels the effect of $f$. So it is too
much to expect that the detection of these processes could make some
constrains  on the scale $f$.
%(v) Finally, with differen c.m. energy
%of the ILC, the cross sections is quite different.

 Note that in our numerical evaluation, only the new particles of the LHT model
  which are not belong to the SM  are concerned. And that, even the
  new particles whose  effects turn out to be of
  high order in $\upsilon/f$ do not enter our analysis as mentioned above.
  Our results show that the cross section of process $e^+e^-\rightarrow b\bar{s}$
  in the LHT model is higher than the value of SM which the cross
  section $\sigma (e^+e^-\rightarrow b\bar{s})$ is of the order of $10^{-3}$
  \cite{eebs-SM}.
%  to be considered in LEP II .

   Based on the above discussion, we know that the LHT model has the
potential to produce large amounts of $b\bar{s}$ events. With the
large yearly luminosity $\mathcal {L}= 500 ~{\rm fb}^{-1}$ and
$\sqrt{s}=500$ GeV at ILC \cite{ILC}, about $10^3$ $b\bar{s}$ events
per year can be produced for case II(2) illustrated in  Fig.5. Such
a signal should appear in the detector as an event with one b-jet
and one light-quark-jet (assuming no distinction is made among light
quarks: $d$, $u$ and $s$). The efficiency to identify b-quark is
about $60\%$ and the signal of one b-quark jet and one
light-quark-jet is certainly FC signal. Although b-quark can be
misidentified as c-quarks, there are still enough $b\bar{s}$ can be
identified. Because the $b\bar{s}$ production is a FC process, the
SM background should be very clean. If a luminosity reaches
$100-1000~fb^{-1}$, the $b\bar{s}$ signal with cross section
$0.1~fb$ could be observed \cite{eebs-SM}. Therefore, with a yearly
luminosity $500~fb^{-1}$ at the ILC, $b\bar{s}$ signal should be
observable if LHT model is right.

\section{Conclusion}

In this paper, we study the processes
$e^+e^-(\gamma\gamma)\rightarrow b\bar{s}$ in the framework of the
LHT model at the ILC. Our results show that the cross sections of
these  processes are strongly dependent on the masses of the third
generation mirror quarks $m_{H_3}$ and the parameterization of the
mixing matrix $V_{H_d}$. And that, the production rate of these
process are sizeable for the favorable values of the parameter.
Hence, it is quite possible to test the LHT model or make some
constrains on the relevant parameters of the LHT through the
detection of theses process.

\section{Acknowledgments}
\vspace{-0.5mm} We would like to thank Junjie Cao and Lei Wu for
helpful discussions and suggestions. This work is supported by the
National Natural Science Foundation of China under Grant
Nos.10775039, 11075045, by Specialized Research Fund for the
Doctoral Program of Higher Education under Grant No.20094104110001,
20104104110001 and by HASTIT under Grant No.2009HASTIT004.

\end{document}